\documentclass[final,authoryear,1p,times]{elsarticle}

\usepackage{amssymb}
\usepackage{epsfig}

\newcommand{\vt}{{\vec{\theta}}}
\newcommand{\vU}{{\vec{ U}}}

\newcommand{\vd}{{\vec{ d}}}

\def\P{{\mathcal{P}}}
\def\V{{\cal V}}

\def\HI{H~{\sc i} \,}

\journal{New Astronomy}

\begin{document}

\begin{frontmatter}

\title{Probing interstellar turbulence in spiral galaxies using HI
  power spectrum analysis}

\author[label1]{Prasun Dutta \corref{cor1}}
\ead{prasun@ncra.tifr.res.in}
\address{ National Centre For Radio Astrophysics, Post Bag 3,
  Ganeshkhind, Pune 411 007, India.} 

\author[label2]{Ayesha Begum}
\ead{begum@astro.wisc.edu}
\address{IISER-Bhopal,ITI Campus (Gas Rahat)Building,Govindpura,
  Bhopal - 23, India}   


\author[label3]{Somnath Bharadwaj}
\ead{somnath@cts.iitkgp.ernet.in}
\address{Department of Physics and Meteorology \& Centre for Theoretical
  Studies, IIT Kharagpur, 721 302  India} 


\author[label1]{Jayaram N. Chengalur}
\ead{chengalu@ncra.tifr.res.in}
\address{ National Centre For Radio Astrophysics, Post Bag 3,
  Ganeshkhind, Pune 411 007, India.}

\begin{abstract}
We estimate the \HI intensity fluctuation power spectrum for a sample of
18 spiral galaxies chosen from  THINGS. Our analysis spans a large
range of length-scales from $\sim 300 \, {\rm pc}$ to $\sim 16 \, {\rm
  kpc}$ across the entire galaxy sample.  We find that 
the power spectrum of each galaxy  can be well fitted by a  power law
$P_{\rm HI}(U) = A\ U^{\alpha}$,  with  an index $\alpha$ that varies
from galaxy to galaxy. For some of the galaxies the  scale-invariant
power-law  power  spectrum  extends to length-scales 
that are comparable to the size of the galaxy's disk. 
The distribution of $\alpha$ is strongly peaked 
with $50 \%$ of the values in the range $\alpha=-1.9$ to $1.5$, and 
a mean  and  standard   deviation of  $-1.3$ and 
$0.5$ respectively. 
We find no significant correlation between 
$\alpha$ and  the star formation rate, dynamical mass,
\HI mass or velocity dispersion of the galaxies. 

Several earlier studies that have measured the power spectrum within
our Galaxy on length-scales that are considerably smaller than $500 \,
{\rm pc}$  have  found  a power-law power spectrum with $\alpha$ in
the range $\approx -2.8$ to $-2.5$. We propose a picture where we
interpret the values in the range $\approx -2.8$ to $-2.5$ as arising
from three dimensional (3D) turbulence in the Interstellar Medium
(ISM) on length-scales smaller than the galaxy's scale-height, and we
interpret the values in the range $\approx -1.9$ to $-1.5$ measured
in this paper  as arising from two-dimensional ISM turbulence in the
plane of the galaxy's disk.  It however still remains a difficulty to
explain the small galaxy to galaxy variations in the values of
$\alpha$ measured here. 
\end{abstract}


\end{frontmatter}

\section{Introduction}
Scale-invariant  structures seen in galaxies are believed to be the
result of  compressible turbulence. The source of the turbulence could
be protostellar winds, supernova, rotational shear, spiral arm shocks,
etc. (see \citealt{2004ARA&A..42..211E} for a review). A variety of
statistical measures have been proposed to characterize the structures
seen in the ISM (see \citealt{2000ApJ...537..720L,
 2001ApJ...555L..33P, 2002ApJ...566..289B, 2004ApJ...616..943L,
2010ApJ...708.1204B}). Of these, power spectrum is most popular.
In the case of compressible  turbulence   the power spectrum is expected
to have a power law shape, and the slope $\alpha$ of the power law contains
information as to the nature of the turbulence.

In the case of the galactic neutral hydrogen (\HI) power spectrum analysis
was first done by \citet{1983A&A...122..282C} who found that the
slope $\alpha$ of the power is $\sim -2.7$ for spatial scales of $\sim
5$ pc 
to $\sim 100$ pc. Similar results were found by
\citet{1993MNRAS.262..327G} for slightly  larger length scales,
i.e. $\sim 200$ pc. The very small scales have been probed using
absorption studies against extended sources. Power law slopes 
in the range $\sim -2.5$ to  $-2.8$ have been found on scales of $0.01$ pc to
$3.0$ pc by \citet{2000ApJ...543..227D} and \citet{2010MNRAS.404L..45R}. Studies of the \HI
in the LMC \citep{1999MNRAS.302..417S} and SMC \citep{2001ApJ...548..749E}
also show that the intensity fluctuations have a power law spectrum.

 Recently, \citet{2006MNRAS.372L..33B} [henceforth Paper I] presented
a visibility-based formalism for determining the power spectrum of external
galaxies with extremely weak \HI emission. Using this formalism they
have found that the power spectrum of the nearby faint  (M$_{\rm B}
\sim -10.9$) dwarf galaxy 
DDO~210 is a power law, i.e, $P(U) = A\ U^{\alpha}$ with  $\alpha \sim
-2.6$. This is consistent with the values of $\alpha$ measured for
our Galaxy and  the nearby galaxies like LMC and  SMC.  In a
subsequent paper \citep{2008MNRAS.384L..34D}  [henceforth Paper II] we
have  used the same visibility based formalism to measure the \HI 
power spectrum  of the  spiral galaxy NGC~628.  The power spectrum was
found to be a power law (on scales of $0.8$ to $8$ kpc), but the slope
$\alpha$ was found to be $\sim −1.6$, i.e.  much flatter than that
determined in the earlier studies. The earlier  studies  all probed
much smaller length scales ($\leq 500$ pc).
\citet{2008MNRAS.384L..34D} proposed that the difference 
arises because at large scales the measurements corresponds to two 
dimensional (2D) turbulence in the plane of the galaxy's disk 
whereas the earlier observations were all restricted to length scales
smaller than the scale height (or thickness) of the galaxy's disk
where three dimensional (3D) turbulence is expected. This picture was
subsequently verified   \citep{2009MNRAS.397L..60D} [henceforth  Paper
  III] in the galaxy NGC~1058  whose \HI power spectrum was found to
be well described by a broken power  law, with slope
$\alpha=-1.0 \pm 0.2$ at large length scales ($1.5$ to $10.0$ kpc)
and another with slope $\alpha=-2.5\pm 0.6$ at small length scales
($0.6$ to $1.5$ kpc).  The transition, which occurs  at the  length
scale $1.5$ kpc, was used to  estimate  the  scale height of the
galaxy's \HI disk as $490 \pm 90$ pc. In a more recent study,
\cite{2010ApJ...718L...1B}   have observed  a similar break in the
power spectrum determined from  {\em Spitzer} observations of the LMC
and  measured the line of sight thickness of LMC to be in the  range
$100-200$ pc. In \citet{2009MNRAS.398..887D} [henceforth paper IV] we
have extended our study of the \HI power spectrum to a sample of five
dwarf galaxies. Paper IV also presents simulations which substantiates
our interpretation of the two different slopes in terms of 2D and 3D
turbulence respectively. In \citet{2010MNRAS.405L.102D}  [henceforth
  Paper V] we have studied the \HI power spectrum of      a harassed
galaxy from the Virgo cluster, where the power spectrum  slope of the
outer part is different from that of the inner part  consistent with
harassement being dominant at the outer parts.
  
In this paper we present the power spectrum of \HI intensity
fluctuations of  $18$ spiral galaxies drawn from the 
THINGS sample. THINGS, “The HI Nearby Galaxy Survey”, is an \HI 21-cm
emission survey of 34 nearby galaxies carried out 
using the NRAO Very Large Array (VLA)
\citep{2008AJ....136.2563W}\footnote{We are indebted to Fabian Walter 
for providing us with the  calibrated \HI data from the THINGS
survey.} with an aim to 
investigate the nature of the interstellar medium 
(ISM), galaxy morphology, star formation and mass distribution across
the Hubble sequence. 
\citet{2008AJ....136.2648D}  present high
resolution rotation curves for the galaxies in the THINGS sample. Star
formation rate and efficiency of the THINGS galaxies are also 
extensively studied in \citet{2008AJ....136.2846B, 2008AJ....136.2782L}.  
This is  as an ideal data set for a comparative study of  the scale invariant 
fluctuations observed  in the neutral ISM of  spiral galaxies. 
 The rest of the paper is organized as
follows. In Section~\ref{sec:data} we discuss  the galaxies in our
sample and briefly mention the power spectrum estimator used in this
study. The results are  presented and discussed in
Section~\ref{sec:results}. Finally we conclude with our main results
in the Section~\ref{sec:conc}. 

\section{Data and Analysis}
\label{sec:data}

\begin{table*}
\centering
\caption{Some parameters of the galaxies used for the power spectrum analysis.
 Columns 1-9 gives 1) Name of the galaxy, 2) and 3) Major and Minor
 axis at a column density of $10^{19}$ atoms cm$^{-2}$, 4) Radius
 corresponding to major axis in kpc, 5) Distance to
 the galaxy, 6) average HI inclination angle,
 7) Star Formation Rate,  8) HI mass, 9)
 Dynamical mass. References are as follows:
(1) \citet{2008AJ....136.2563W},  
(2) \citet{2008AJ....136.2648D},  
(3) \citet{1973A&A....29..425B},  
(4) \citet{1979A&A....74..138H},  
(5) \citet{1985A&A...143..216H},  
(6) \citet{1992A&A...253..335K},  
(7) \citet{1996ApJ...458..120S}.  
}
\begin{tabular}{l c c c c c c c c c c|}
\hline \hline
Galname & Major & Minor & R$_{maj}$ & D & $i_{HI}$ & log(SFR) & $M_{HI}$ &  $M_{dy}$ &  References\\ 
  & $(')$ & $(')$ & (kpc) & (Mpc) & $(^{\circ})$ & (M$_{\odot}$ yr$^{-1}$) & ($10^{8}$ M$_{\odot}$) & ($10^{11}$ M$_{\odot}$) & \\ 
\hline  
 NGC~628  & $22.0$ & $20.0$ & $23.4$ & $ 7.3$ & $ 15.0$ & $1.2$ & $38.8$ & $6.3$  & 1,2,6\\ 
 NGC~925  & $16.0$ & $10.0$ & $21.4$ & $ 9.2$ & $50.0$ & $1.1$ & $45.8$ & $ 1.7$  & 1,2\\ 
 NGC~2403 & $25.0$ & $22.0$ & $11.6$ & $ 3.2$ & $55.0$ & $0.9$ & $25.8$ & $ 3.1$  & 1,2\\ 
 NGC~2841 & $26.0$ & $22.0$ & $53.3$ & $14.1$ & $69.0$ & $0.2$ & $85.8$ & $31.2$  & 1,2\\ 
 NGC~2903 & $25.0$ & $15.0$ & $32.4$ & $ 8.9$ & $66.0$ & $  -$ & $43.5$ & $ 4.3$  & 1,2\\ 
 NGC~3031 & $38.0$ & $24.0$ & $19.9$ & $ 3.6$ & $59.0$ & $1.1$ & $36.4$ & $ 5.2$  & 1,2\\ 
 NGC~3184 & $17.0$ & $14.0$ & $27.4$ & $11.1$ & $29.0$ & $1.4$ & $30.7$ & $ 6.4$  & 1,2,5\\ 
 NGC~3198 & $22.0$ & $ 7.5$ & $41.6$ & $13.8$ & $72.0$ & $0.9$ & $101.7$& $ 8.1$  & 1,2\\ 
 NGC~3521 & $22.0$ & $ 7.5$ & $34.2$ & $10.7$ & $69.0$ & $8.4$ & $80.2$ & $12.2$  & 1,2\\ 
 NGC~3621 & $25.0$ & $15.0$ & $24.0$ & $ 6.6$ & $62.0$ & $2.1$ & $70.7$ & $12.8$  & 1,2\\ 
 NGC~4736 & $18.0$ & $12.0$ & $12.3$ & $ 4.7$ & $44.0$ & $0.4$ & $ 4.0$ & $ 1.2$  & 1,2\\ 
 NGC~5055 & $30.0$ & $25.0$ & $44.1$ & $10.1$ & $51.0$ & $2.4$ & $91.0$ & $12.8$  & 1,2\\ 
 NGC~5194 & $16.0$ & $12.0$ & $18.6$ & $ 8.0$ & $30.0$ & $6.1$ & $25.4$ & $2.7$   & 1,2,7\\ 
 NGC~5236 & $30.0$ & $24.0$ & $19.6$ & $ 4.5$ & $31.0$ & $2.5$ & $17.0$ & $2.6$   & 1,2,3\\ 
 NGC~5457 & $30.0$ & $25.0$ & $32.3$ & $ 7.4$ & $30.0$ & $2.5$ & $141.7$& $5.9$   & 1,2,4\\ 
 NGC~6946 & $35.0$ & $25.0$ & $30.0$ & $ 5.9$ & $35.0$ & $4.8$ & $41.6$ & $ 7.3$  & 1,2\\ 
 NGC~7793 & $12.0$ & $ 9.0$ & $6.8$ & $ 3.9$ & $43.0$ & $0.5$ & $ 8.9$ & $ 0.7$  & 1,2\\ 
 IC~2574  & $14.0$ & $ 8.0$ & $8.1$ & $ 4.0$ & $51.0$ & $0.1$ & $14.8$ & $0.5$   & 1,2\\ 
\hline
\end{tabular}

\label{tab:sample}
\end{table*}

THINGS provides  high angular ($\sim$6$^{\prime\prime}$)  and velocity
($\leq$ 5.2 kms$^{-1}$) resolution   VLA observations of a sample of
34 nearby galaxies spanning a range of Hubble types, from  dwarf 
irregulars to massive spirals.  In this paper our analysis is restricted
to spiral galaxies  with minor axis greater than $6'$.
 Table~\ref{tab:sample} lists different properties of the
galaxy sub-sample that we have analyzed. 
The different columns in  Table~\ref{tab:sample}   are as follows: 
column (1) Name of the galaxy, (2) and (3)the \HI major and minor axis
respectively at a column density of $10^{19}$ atoms cm$^{-2}$, (4)
radius corresponding to the major axis in kpc, (5) distance to the
galaxy, (6) the average \HI inclination angle, 
(7) the star formation rate, (8) the \HI mass and (9) the dynamical
mass derived from the rotation curve. The values for the following parameters:
distance to the galaxy, 
star formation rate and the total \HI mass are taken from
\citet{2008AJ....136.2563W}, 
whereas values for the inclination angles are adopted from
\citet{2008AJ....136.2648D}. 
\cite{2008AJ....136.2648D} presents rotation curves for $13$ 
galaxies in our sample derived from the same data as the ones that we are using
for the current analysis. We use these rotation curves to estimate
the dynamical mass of these galaxies. For the rest of the
galaxies, the values of the dynamical mass  are taken from
\citet{1973A&A....29..425B, 1979A&A....74..138H, 1985A&A...143..216H,
  1992A&A...253..335K, 1996ApJ...458..120S}  and then   
rescaled for the adopted distances noted in Table~\ref{tab:sample}.
 Though the galaxy NGC~4826 satisfies 
our selection criteria,  we have not used it in our analysis for reasons
which are detailed later.
 
For our analysis, we  started with the calibrated visibility  data prior to
continuum subtraction.
In the standard THINGS pipeline, the continuum is subtracted by fitting a
linear polynomial to each visibility (i.e. AIPS task UVLIN). This
method is computationally fast and works very well in many 
cases. However, any extended source, (or even off axis point sources) would 
have visibilities that oscillate with frequency and are not well fit
by a straight line. We hence model (using clean components) the continuum 
emission in the field using the line free channels. This contribution 
of the model continuum to the visibilities at all frequencies was
then estimated and subtracted out using the AIPS task UVSUB.
 The resulting continuum-subtracted data was used for the subsequent 
 analysis. We have identified the spectral channels that  contain 
 \HI emission  with a relatively high  signal to
noise ratio, and used only these channels for the power spectrum analysis. 

We have tested the efficacy of our  continuum subtraction by comparing the 
angular power spectrum  of the channels with \HI emission with the the angular
power spectrum of the line-free channels. The line-free channels give
an estimate 
of the contributions from the residuals after continuum subtraction.
Our results,  
shown later in this paper, clearly demonstrate that the \HI signal is
significantly in   
excess of the  residuals from continuum subtraction. The subsequent analysis
is entirely restricted to the range where the \HI signal dominates
over the residual 
 continuum,   
and hence we do not expect the contribution from the residual
continuum to make a  
significant contribution to the estimated \HI power spectrum. 

\subsection{The \HI power spectrum estimator}

The HI power spectrum can be estimated either from the deconvolved
data cubes or the calibrated visibilities. The advantage of using
the calibrated visibilities is that one is dealing directly with
the measured quantities whose noise properties are well understood.
On the other hand most popularly used deconvolutions involve
heuristics, and the  noise properties of the final deconvolved data cubes are
not well understood. Previous works that used visibility based
estimators of the power spectrum include \citet{1983A&A...122..282C,
1993MNRAS.262..327G}, and \citet{1995A&A...293..507L}. Our method differs
from these in the way in which the noise bias is handled. The noise
bias may be neglected in observations of Galactic HI where the
signal to noise ratio is large. In observations of external galaxies
where the signal to noise ratio is modest, it is important to handle
the noise bias properly. Paper I and Paper IV together provide a
detailed description of the 
technique that we use to estimate the power spectrum. However, for the
convenience of the reader, we briefly present the salient features below.  

We model the \HI brightness distribution of  an external galaxy as  
\begin{equation}
I (\vt)=W(\vt)~[\bar{I} + \delta I(\vt)],
\label{eq:a0}
\end{equation}
where we assume that the angular extent 
of the galaxy is sufficiently  small so that we may represent angular separations 
as two dimensional vectors $\vt$ in the sky plane.   The term 
$W(\vt) \bar{I}$ describes the global \HI distribution of the galaxy
{\it ie.} the smooth fall off of brightness with
 angular distance away from the center of the galaxy. This 
can be approximately modeled as an exponential
\begin{equation}
 W(\vt)=\exp(\sqrt{12} \theta/\theta_0)
\end{equation}
 for face-on disk galaxies (see Paper IV) where $\theta_0$ corresponds to the 
angular extent of the galaxy's \HI disk. We conservatively use the
angular extent of the semi-minor axis for  $\theta_0$ when the disk is
inclined to the line of sight.  
  The term  $\delta I(\vt)$ represents 
the fluctuations in the \HI brightness. These fluctuations are modulated by 
the galaxy's global \HI profile $W(\vt)$ which we can think of as a window function.  We 
assume that the 
fluctuations  $\delta I(\vt)$ are the outcome of homogeneous and isotropic
compressible turbulence,  and hence  $\delta I(\vt)$ is  a 
statistically homogeneous and isotropic random field. It is important to note that the 
assumption regarding the source of the fluctuations ({\it ie.} turbulence) 
is not crucial in our interpretation, it could equally well be some other 
mechanism. However, the assumption that fluctuation $\delta I(\vt)$ is
 statistically homogeneous and isotropic plays a crucial role in our analysis. 
The assumption that   $\delta I(\vt)$ is statistically homogeneous 
implies that the two-point correlation function $\xi$ defined as 
\begin{equation}
\xi(\vt_1,\vt_2) = \langle \delta I(\vt_1)\, \delta I(\vt_2) \rangle.
\label{eq:a1}
\end{equation}
depends only only on the separation $\vt_1-\vt_2$ 
 {\it ie.} $\xi(\vt_1,\vt_2) \equiv \xi(\vt_1-\vt_2)$. The assumption of statistical 
isotropy further implies that $\xi$ only depends on the 
magnitude of the angular separation $\mid \vt_1-\vt_2 \mid$ {\it ie.} 
$\xi(\vt_1,\vt_2) \equiv \xi(\mid \vt_1-\vt_2 \mid)$. The angular brackets in eq.
 (\ref{eq:a1} ) denotes
an average over different realizations of the random field $\delta I(l,m)$. 

The angular power spectrum $P_{\rm{HI}}(\vU)$ of the \HI brightness fluctuations  
 is the quantity of interest in this paper. The power spectrum $P_{\rm{HI}}(\vU)$ is
the  Fourier transform of the two-point correlation function
\begin{equation}
P_{\rm{HI}}(\vU)= \int~\xi(\vt)~
e^{- i 2 \pi \vU \cdot \vt   }~d^2 \theta.
\label{eq:a2}
\end{equation}
where $\vU$ is a two-dimensional vector that refers to inverse angular scales. 
The assumptions of statistical homogeneity and isotropy imply that $\xi(\vt)$ and 
$P_{\rm{HI}}(\vU)$ depend only on $\theta$ and $U$,  the magnitudes  of the respective
vectors. 

Now radio-interferometric observations directly measure a quantity related
to $P_{\rm{HI}}(\vU)$, viz. the visibility,
\begin{equation}
{\cal V}(\vU)=\int I(\vt)\ e^{- i 2 \pi \vU \cdot \vt   }~d^2 \theta.
\label{eq:a3}
\end{equation}
which is the  Fourier  transform of the brightness distribution.
Here the two-dimensional vectors $\vU$, referred to as baselines, denote the
 separation $\vd$ between pairs of antennas  measured in units of the 
observing wavelength $\lambda$ ({\it ie.} $\vU=\vd/\lambda$). The antenna separations
$\vd$ are projected onto the plane normal to direction of observation,
{\em i.e} a plane parallel to the sky plane.

We return to the relation between $\vU$ and $P_{\rm{HI}}(\vU)$ below,
but first we  discuss the option of using eq.~(\ref{eq:a3}) to make a image of
the source, and then determine $P_{\rm{HI}}(\vU)$ from this image. As
is well known,  it is possible to invert the Fourier relation
(eq.~\ref{eq:a3}),   and use the  
measured visibilities to reconstruct the sky brightness distribution $I(\vt)$. 
The visibilities are typically not available for all the $\vU$ values that are
required to reconstruct the galaxy's image.  Consequently, image reconstruction
involves complicated  non-linear and non-local deconvolution techniques which 
rely on human judgment to a certain extent. Deconvolution is particularly 
important in our context where we are not interested in any particular 
visible feature in the galaxy’s image. On the contrary, we are intent 
on quantifying the statistical properties of the random field 
$\delta I(\theta)$. The effect of the deconvolution on the $\delta I(\theta)$
is difficult to quantify. Further, the noise in the different pixels 
of deconvolved image are correlated - the actual correlation properties
are also difficult to quantify. For these reasons, it is non trivial
to quantify the uncertainties in the power spectrum obtained from
the deconvolved images. These problems can be avoided if one directly
works with the visibilities, where the noise properties are relatively
straight forward. We describe such an estimator below. We note in
passing, that although we do not use the image to estimate the power
spectrum, we have  made images of all the galaxies that we analyze.
These images were only used to estimate the angular extent of the HI 
disk and to determine the channels with significant HI emission.

The most straight forward visibility based power spectrum estimator
\citep{1983A&A...122..282C} is the square of the visibilities ($\mid
      {\cal V}(\vU)\mid^2$). However, in addition to the \HI signal from the 
galaxy each visibility also has a noise contribution. The latter introduces a 
positive noise bias in the power spectrum estimator $\mid {\cal V}(\vU)\mid^2$. 
While it is, in principle, possible to separately estimate the noise bias and 
subtract it out, this is extremely difficult in practice. The noise bias is 
often larger than the \HI power spectrum, and the uncertainties in calibration 
and  noise estimates make it extremely difficult to reliably remove the noise bias. 

Paper I presents a technique to estimate the \HI power spectrum, avoiding the problem
of noise bias.  This technique is based on the fact that the  visibilities  at the different
baselines within a disk of radius $(\pi \theta_{0})^{-1}$ around a baseline $\vU$ are all
correlated and   they can all be used to estimate the power spectrum $P_{\rm HI}(U)$. 
We correlate the visibility $\V(\vU)$ with all other visibilities  
$\V(\vU+\Delta \vU)$  within a disk $|\Delta \vU| < (\pi \theta_{0})^{-1}$
to obtain the estimator $\P_{\rm HI}(U)=\langle \V(\vU) \, \V^{*}(\vU+\Delta \vU) \rangle$. 
The self correlation of a visibility with itself is not included to  avoided the 
noise bias.  The estimator is further averaged over different directions $\vU$ and 
binned to increase the signal to noise ratio. The relationship between
the estimator $\P_{\rm HI}(U)$ and the \HI power spectrum $P_{\rm
  HI}(U)$ can be shown to be (see paper IV) 
\begin{equation}
\P_{\rm HI}(\vU)=
\mid {\tilde W}(\vec U)\mid^2 ~\otimes~P_{\rm HI}(\vU) \, ,
\label{eqn:first}
\end{equation}
where ${\tilde W}(\vec U)$ is the Fourier transform of the window 
function $W(\vt)$ introduced in eq. (\ref{eq:a0}). 
For a window function centered at $\theta =0$ and with angular extent 
$\theta_{0}$, we expect  $\mid \tilde W(\vec U)\mid^2$ to be 
sharply peaked at $\vec U=0$ and to fall  off rapidly for $U \gg \theta_{0}^{-1} $.
Hence, at large baselines, it is adequate to approximate $\mid \tilde W(\vec U)\mid^2$ 
by a Dirac delta function in eq. (\ref{eqn:first}), whereby 
\begin{equation}
\P_{\rm HI}(\vU)=  K \, P_{\rm HI}(\vU)
\label{eq:b1}
\end{equation}
 {\it ie.}  we may ignore the convolution and   $\P_{\rm HI}(U)$ gives a
 direct estimate 
of the \HI  power spectrum   $ P_{\rm HI}(U)$.  Numerical studies  (Paper IV) 
show that it is possible to identify a baseline $U_m$ such that eq. (\ref{eq:b1}) 
holds for all $U > U_m$, and we may directly interpret  the 
estimator $\P_{\rm HI}(\vU)$ 
as the \HI power spectrum $P_{\rm HI}(\vU)$ at baselines larger than
$U_m$. At baselines smaller than $U_{m}$, the window function has a
non-negligible effect on the slope of the power spectrum estimator.
 The value of $U_m$ is proportional to $\theta_0^{-1}$, and it also depends on the 
slope of the \HI power spectrum (Paper IV).  In our analysis, we have separately
determined $U_m$ for each galaxy in our sample and used only the range $U > U_m$ 
for the subsequent analysis. It should be noted that the estimator $\P_{\rm HI}(U)$
has both real and imaginary parts, and we use only the real part to estimate the 
\HI power spectrum. We expect the imaginary part to be zero if  
 the various assumptions made in our analysis  are all perfectly valid. The measured 
estimator usually has a small  imaginary component. This arises due to the limitation of the 
various assumptions and also as a consequence of noise. The requirement that the real part 
of the estimator should be considerably larger than the imaginary part serves as 
an useful check of our method.  In the subsequent analysis 
we have also  used this requirement to constrain the $U$ range where
it is possible to  
make a reliable estimate of the \HI power spectrum.

For each galaxy in our sample, we determine  if the estimated power spectrum can be
modeled as a power law $P_{{\rm HI}}(U) = A~U^{\alpha}$.  The largest baseline
$U_{max}$  used while doing the  power law fit is determined by the criteria that the real 
part of the estimator should be greater than the imaginary part, and also greater
than the power spectrum estimated from  the line-free channels. 
We first make an initial estimate  of the smallest baseline $U_m$ 
from a visual inspection of the estimated power spectrum. 
We use the range $U_m \le U \le U_{max}$ to  fit a power law  
through a $\chi^2$    minimization. The 
best fit $\alpha$ obtained by this fitting procedure  and the galaxy's
angular extent 
are both used to  get a revised  estimate for $U_m$  (see 
  Figure~4 in paper IV).  
We iterate through these steps a few times till it converges, and   
this gives us a range of  baseline ($U_{min}$ to $U_{max}$) over which
we have a power law fit to the power spectrum. To test  that the 
window function  actually has a very small  effect,  we convolve the best fit
power law  with $\mid  \tilde{W}(U)\mid^2$ assuming an exponential
window function.  The goodness of
fit to the data is also estimated by calculating $\chi^2$ for the
convolved power spectrum.  We accept the final fit only after
ensuring that the effect of the convolution can actually be ignored. 
The method that we have used  to determine the best fit power law 
is only sensitive to the  slope of the power spectrum  and not 
the  amplitude.  A detailed model of the window function (as opposed
to the simple exponential assumed here) is needed 
to determine the amplitude of the \HI power spectrum, and we have not attempted this
in the present paper.  Finally, we have also converted the 
range of baselines $(U_{min},U_{max})$ to a range of length-scales 
$(R_{max},R_{min})$ using $R=D/U$ where   $D$ is the 
distance to the galaxy. This  provides an estimate of  the range of
length-scales where the power-law fit holds. 

Fluctuations in the \HI specific  intensity could arise from
fluctuations in either  the density or  the velocity. 
The velocity fluctuations, however, are not important 
if the width of the  frequency slice is larger  than the galaxy's
velocity 
dispersion  \citep{2000ApJ...537..720L}. In our analysis we have 
first averaged the visibilities over all the frequency 
channels that contain  significant  \HI emission,  and  we then use
these   to estimate the power spectrum. We have checked that the
resulting channel width $\delta v$ is larger than the typical velocity
dispersion for all the galaxies in our sample. The channel averaging
effectively collapses the signal, and we may interpret the measured
power spectrum as being proportional to the angular power spectrum of
the fluctuations in the \HI column density.

The $1\sigma$ error-bars for the estimated power spectrum are a sum,
in quadrature,  of the contributions from two different sources of uncertainty. 
At small $U$ (i.e. large angular scales) the uncertainty is dominated 
by the sample variance which comes from the fact that we have a finite
and limited number of independent estimates of the true power spectrum. 
At large $U$ (i.e. small angular scales), it is dominated by the
system noise in the visibilities. We have used these error bars in our
$\chi^2$ minimization procedure. Details of the error estimation can
be found in \citet{2011arXiv1102.4419D}.

 As mentioned earlier,  the galaxy NGC~4826 meets all the requirements of our selection 
criteria. However, this galaxy  has a very bright \HI core which makes 
the window function$W(\vt)$  complicated,  and  it is not possible to
conclusively determine $U_{m}$ for this galaxy. We hence not include 
this galaxy in our sample. 

\section{Result and Discussion}
\label{sec:results}

\begin{table*}
\centering
\caption{Result of the power spectrum analysis.  Column 1 to 7 gives
  1)   name of the galaxy, 2) width of the 
  channel used to estimate the   power spectrum, 3) and 4) the range
  of $U$ value for which the    power spectrum is evaluated, 5) and 6)
  correspoindin length scales   and 7) the power law index $\alpha$
  with $1-\sigma$ error} 
\begin{tabular}{ l r r r r r r r }
\hline \hline
Galaxy & & $\Delta \ v$ & $U_{min}$   & $U_{max}$     & $R_{min}$ & $R_{max}$ & $\alpha$ \\ 
       & & ($km\ s^{-1}$) & (k $\lambda$) &  (k $\lambda$) & (kpc)      & (kpc)
&   \\ 
\hline 
 NGC~628 & & $41.6$ & $1.0$ & $10.0$ & $0.8$ & $7.5$ & $-1.6 \pm 0.1$ \\ 
 NGC~925 & & $41.6$ & $1.0$ & $10.0$ & $0.9$ & $9.2$ & $-1.0 \pm 0.2$  \\ 
 NGC~2403 &  & $83.2$ & $0.7$ & $7.0$ & $0.6$ & $4.0$ & $-1.1 \pm 0.1$ \\ 
 NGC~2841N & & $83.2$ & $1.0$ & $10.0$ & $1.4$ & $14.0$ & $-1.7 \pm
 0.2$  \\ 
 NGC~2841S & & $83.2$ & $1.0$ & $10.0$ & $1.4$ & $14.0$ & $-1.5 \pm
 0.2$  \\ 
 NGC~2903 & & $83.2$ & $0.8$ & $8.0$ & $1.1$ & $11.1$ & $-1.5 \pm 0.2$ \\ 
 NGC~3031N & & $41.6$ & $2.0$ & $10.0$ & $0.4$ & $1.8$ & $-0.7 \pm 0.1$\\ 
 NGC~3184 & & $41.6$ & $0.7$ & $7.0$ & $1.6$ & $15.8$ & $-1.3 \pm 0.2$ \\ 
 NGC~3198 & & $83.2$ & $1.6$ & $10.0$ & $1.4$ & $8.6$ & $-0.4 \pm
 0.3$  \\ 
 NGC~3521N & & $83.2$ & $1.0$ & $17.0$ & $0.6$ & $10.7$ & $-1.8 \pm 0.1$ \\ 
 NGC~3521S & & $83.2$ & $1.0$ & $17.0$ & $0.6$ & $10.7$ & $-1.6 \pm
 0.2$ \\ 
 NGC~3621 & & $83.2$ & $1.0$ & $12.0$ & $0.6$ & $6.6$ & $-0.8 \pm 0.2$ \\ 
 NGC~4736 & & $83.2$ & $0.6$ & $10.0$ & $0.5$ & $7.8$ & $-0.3 \pm 0.2$\\ 
 NGC~5055 & & $83.2$ & $1.0$ & $10.0$ & $1.0$ & $10.0$ & $-1.6 \pm 0.1$ \\ 
 NGC~5194 & & $83.2$ & $1.0$ & $8.0$ & $1.0$ & $8.0$ & $-1.7 \pm 0.2$  \\ 
 NGC~5236 & & $83.2$ & $0.6$ & $6.0$ & $0.8$ & $7.5$ & $-1.9 \pm 0.2$ \\ 
 NGC~5457 & & $83.2$ & $0.6$ & $12.0$ & $0.6$ & $12.3$ & $-2.2 \pm
 0.1$ \\ 
 NGC~6946 & & $20.8$ & $1.5$ & $10.0$ & $0.3$ & $4.0$ & $-1.6 \pm 0.1$ \\ 
 NGC~7793 & & $41.6$ & $0.6$ & $6.0$ & $0.6$ & $6.5$ & $-1.7 \pm 0.2$ \\ 
 IC~2574 & & $41.6$ & $1.8$ & $10.0$ & $0.4$ & $3.3$ & $-1.7 \pm 0.3$ \\ 
\hline
\end{tabular}
\label{tab:result}
\end{table*}

\begin{figure}
\begin{center}
\epsfig{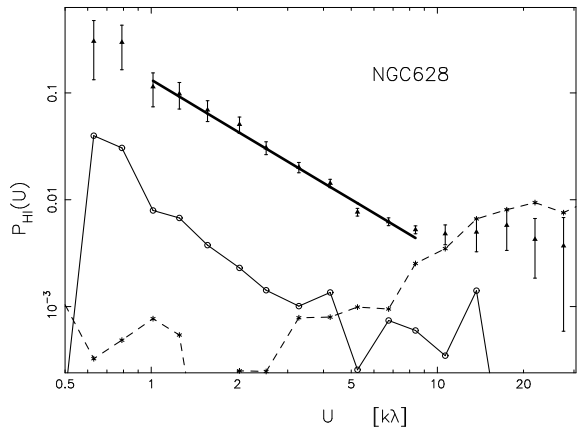}
\end{center}
\caption{The real (filled triangle) and imaginary (open circle) parts of the
  estimator evaluated using the channels with HI emission is plotted
  with their $1\sigma$ error bars for the   galaxy NGC~628. The real
  part for the   line-free channels (star) is also
  shown. The best fit power law is shown  by the straight line. Only
  the baseline range where  the real part   of the estimator  is
  larger than its  imaginary part and  also the real part 
for   the   line-free channels is used to estimate 
 the power spectrum. 
} 
\label{fig:fullU}
\end{figure}

Figure~\ref{fig:fullU} shows the estimated  \HI power spectrum
for a particular  galaxy NGC~628. For this galaxy, we  show the result 
over the  baseline range $0.5$~k$\lambda$ to
 $30$~k$\lambda$, even  though the observational data extends to  larger baselines
($U>30$~k$\lambda$). The \HI signal ($P_{\rm HI}(U)$) falls at  the larger baselines 
which are also sparsely sampled, and  the power spectrum estimated  at these baselines 
is too noisy to be useful. Further,  the  residuals from continuum subtraction 
(Figure~\ref{fig:fullU})  become comparable to the \HI signal
at the large baselines. For channels with \HI emission in
the $U$ range $1.0$~k$\lambda$ to  $10.0$~k$\lambda$, the real part of
the estimator $\P_{\rm  HI}(U)$ is larger than the imaginary part as
well as the power spectrum measured from the line-free channels. We
have restrict the power law fit  to 
the $U$ range  $1.0$~k$\lambda$ to $10.0$~k$\lambda$ (i.e, $U_{min}, \, U_{max}$). 
For the remaining galaxies we only show results for the $U$ range that
is useful for determining  $\P_{\rm  HI}(U)$.

\begin{figure*}
\begin{center}
\epsfig{file=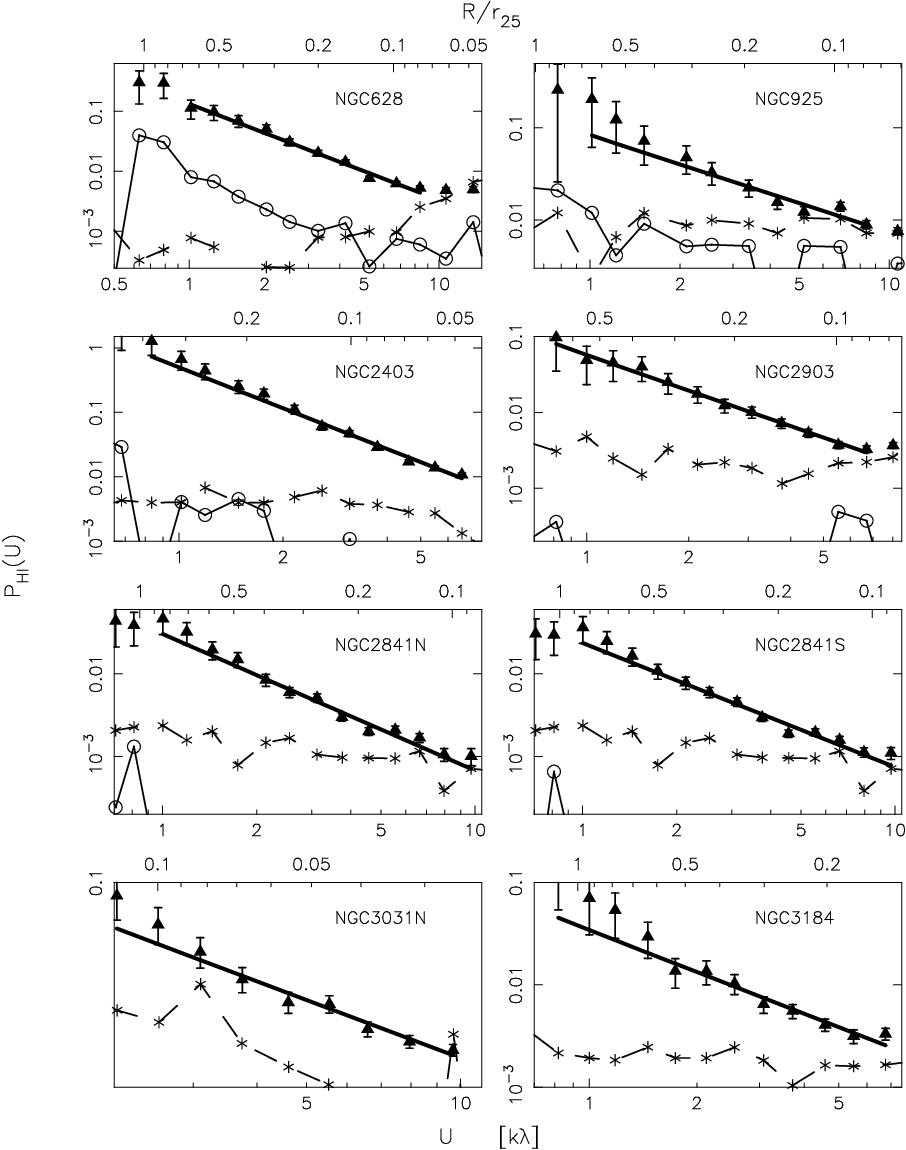, height=5.4in, angle=-90}
\end{center}
\caption{The real (filled triangle) and imaginary (open circle) parts of the
  estimator evaluated using the channels with HI emission is plotted
  with  $1\sigma$ error bars for the galaxy indicated in each of the
  individual panels.    The real
  part for the   line-free channels (star) is also
  shown. The best fit power law is shown  by the straight line. Only
  the baseline range where  the real part   of the estimator  is
  larger than its  imaginary part and  also the real part 
for   the   line-free channels is used to estimate 
 the power spectrum.  The top   margin of the each panel shows the 
 length scale corresponding to the baseline value. The length-scales
 are shown in  units of the  $r_{25}$  of  the   respective galaxy. The
 $r_{25}$ values are adopted from the R3   catalogue of galaxies. }
\label{fig:fig1}
\end{figure*}

\begin{figure*}
\begin{center}
\epsfig{file=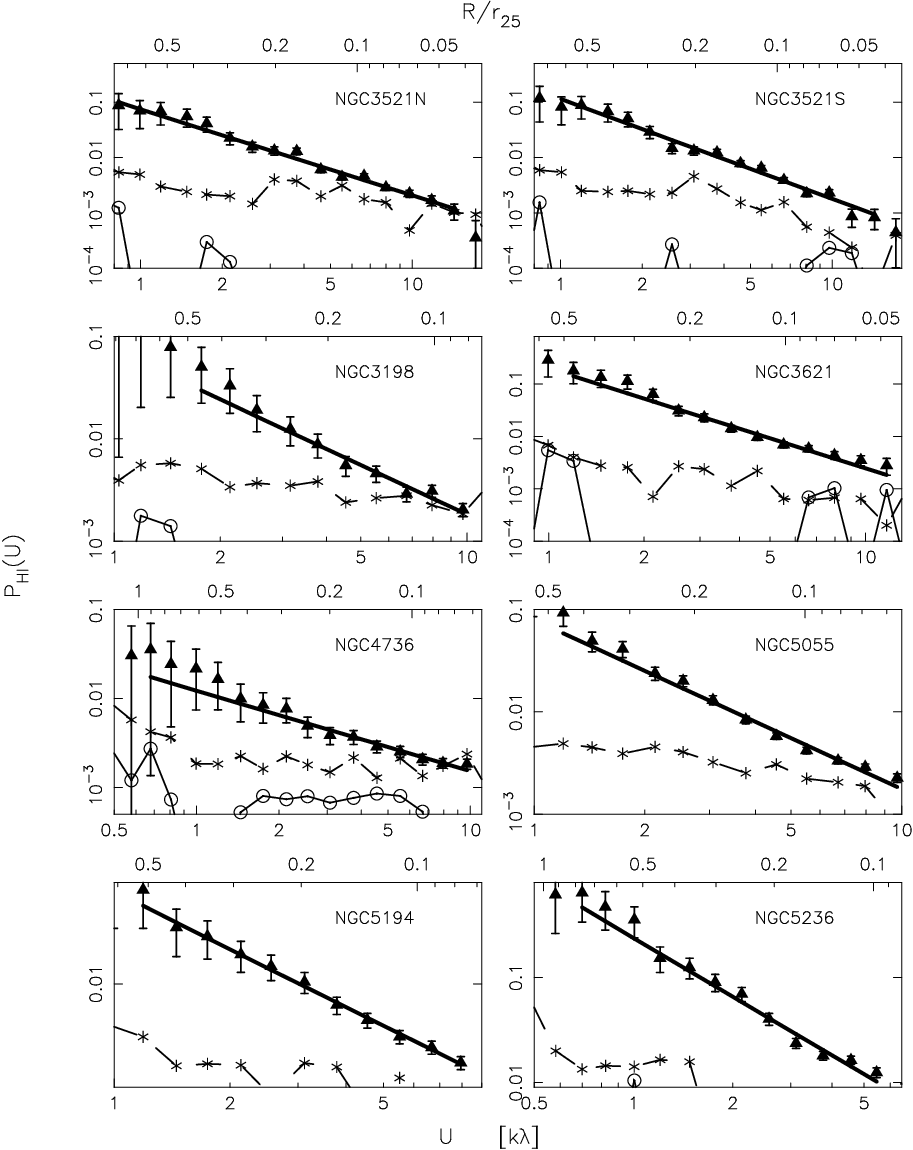, height=5.4in, angle=-90}
\end{center}
\caption{The real (filled triangle) and imaginary (open circle) parts of the
  estimator evaluated using the channels with HI emission is plotted
  with  $1\sigma$ error bars for the galaxy indicated in each of the
  individual panels.    The real
  part for the   line-free channels (star) is also
  shown. The best fit power law is shown  by the straight line. Only
  the baseline range where  the real part   of the estimator  is
  larger than its  imaginary part and  also the real part 
for   the   line-free channels is used to estimate 
 the power spectrum.  The top   margin of the each panel shows the 
 length scale corresponding to the baseline value. The length-scales
 are shown in  units of the  $r_{25}$  of  the   respective galaxy. The
 $r_{25}$ values are adopted from the R3   catalogue of galaxies. }
\label{fig:fig2}
\end{figure*}

\begin{figure*}
\begin{center}
\epsfig{file=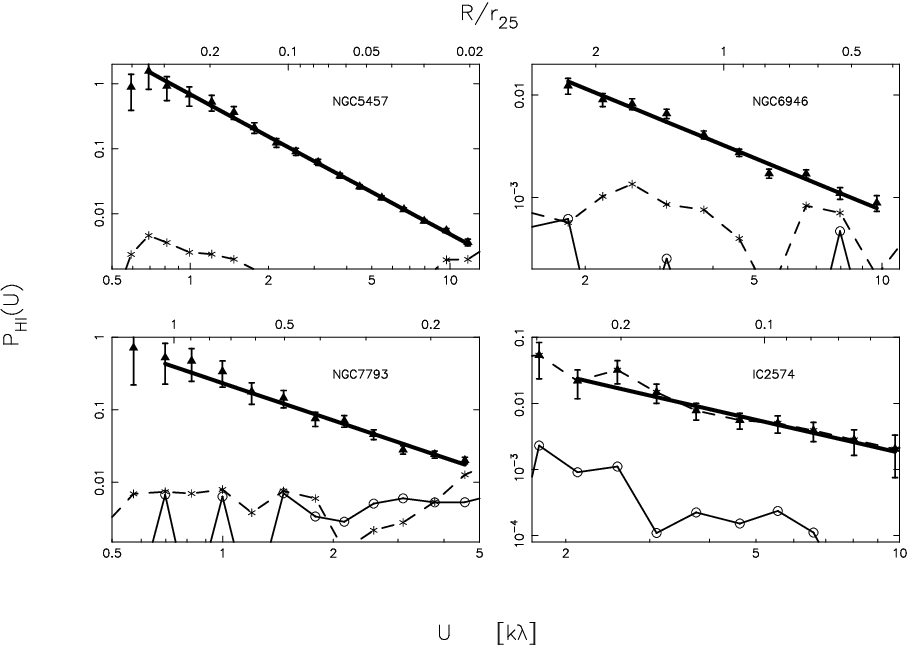, height=5.6in, angle=-90}
\end{center}
\caption{The real (filled triangle) and imaginary (open circle) parts of the
  estimator evaluated using the channels with HI emission is plotted
  with  $1\sigma$ error bars for the galaxy indicated in each of the
  individual panels.    The real
  part for the   line-free channels (star) is also
  shown. The best fit power law is shown  by the straight line. Only
  the baseline range where  the real part   of the estimator  is
  larger than its  imaginary part and  also the real part 
for   the   line-free channels is used to estimate 
 the power spectrum.  The top   margin of the each panel shows the 
 length scale corresponding to the baseline value. The length-scales
 are shown in  units of the  $r_{25}$  of  the   respective galaxy. The
 $r_{25}$ values are adopted from the R3   catalogue of galaxies. }
\label{fig:fig3}
\end{figure*}

 Figures~\ref{fig:fig1},  \ref{fig:fig2} and 
  \ref{fig:fig3} show  the estimated \HI power spectrum  for all the
  galaxies in our 
sample. The results are also summarized in Table~\ref{tab:result}  where
column (2) to (7) gives the velocity width of  the  collapsed channel,
the baseline range used for the fit, the corresponding range of
length-scales and the best fit  power law index $\alpha$
respectively.  

The angular extent  of the galaxies NGC~2841,
NGC~3031 and NGC~3521 are comparable to the telescope's  field of view.  
For each of these galaxies, THINGS contains two distinct 
sets of observations each with a different pointing center   referred to as 
North (N) and South (S) respectively. Each data set has been  separately analyzed
and reported in Table~\ref{tab:result} with the suffix S or N to distinguish between 
them.  The power spectrum could be estimated only over a limited baseline range
 for  NGC~3031S,  and hence we do not consider it here. For both NGC~2841 and NGC~3521,  
we find that the  slope $\alpha$  estimated respectively from the N and the S 
data are consistent with one another.  In the subsequent analysis,  
for these two galaxies we have used the  values of $\alpha$ 
corresponding to the N data. 

In our analysis (Table~\ref{tab:result}), for most of the galaxies  we have been able to 
model the \HI power spectrum using a  power law  $P_{\rm HI}(U)=A U^{\alpha}$  across a 
baseline  range which approximately spans a decade in $U$.  We find that 
NGC~3031N has the smallest dynamical range with $(U_{min},U_{max})=(2,10) \, 
{\rm k} \lambda$. We recollect that this galaxy has two data sets, and the above results
are only from the N data, the \HI power spectrum from the S data could not be fitted with 
a power law over a significant  $U$ range. In addition to this, there are four other 
galaxies (NGC~3198,NGC~5194, NGC~6946 and IC~2574) that have  a dynamical range
less than a decade $(U{max}/U_{min}  < 10)$. The power law fit in NGC~5457 
covers the largest dynamical range with $(U_{min},U_{max})=(0.6,12) \, 
{\rm k} \lambda$, which corresponds to the  range of length-scales 
$0.6$ kpc to $12.3$ kpc. The power spectrum is found to have a slope
$\alpha=-2.2 \pm 0.1$ for this galaxy. This galaxy has the steepest slope 
in our sample, and in fact, it is the only galaxy with  $\alpha
<-2$. We also find, the smallest length-scale ( $0.3$ kpc) occurs in  
NGC~6949 where  $\alpha=-1.6\pm0.1$, and the largest length-scale
($15.8$ kpc)
occurs in NGC~3184  where $\alpha=-1.3\pm0.2$.  Over the range of length-scales
probed in this sample, there does not seem  to be any correlation between the 
length-scale  and the slope $\alpha$. However, 
it is perhaps more relevant to interpret  length-scales in terms to the size of the
galaxy. We have used the value of  $r_{25}$ to 
characterize the intrinsic sizes of the different galaxies
in our sample\footnote{The $r_{25}$ 
values have been  adopted from the R3 catalog of galaxies}.
 For each galaxy, the length-scales $R=D/U$ have been  expressed 
in units of $r_{25}$  and  $R/r_{25}$ has been shown at the top margins of the  plots in 
 Figures~\ref{fig:fig1},  \ref{fig:fig2} and   \ref{fig:fig3}. 
Our results clearly shows the presence of scale invariant 
fluctuations with a power law power spectrum that extends to length-scales comparable to 
the radius of  the galaxy's optical disk ($R/r_{25} \sim 1$). In fact,
we are able to fit a  power law  all the way to $R/r_{25} >2$  in
NGC~6946. We note that $r_{25}$ is a measure of the optical disk size,
the \HI disk size is generally somewhat larger.


\begin{figure}
\begin{center}
\epsfig{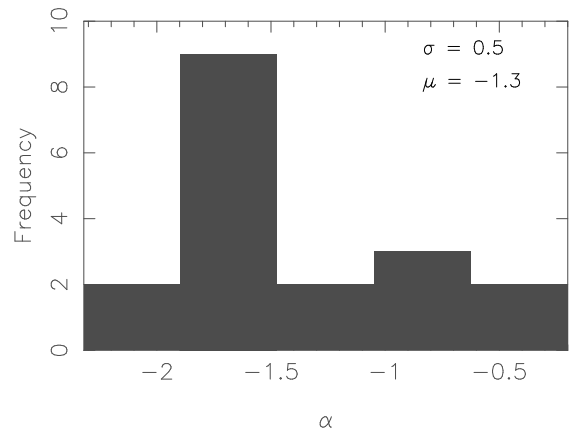}
\end{center}
\caption{Shaded regions shows the histogram of the distribution of the
power law index $\alpha$. Note that the histogram has 5 bins for 18
data points. The sample mean $\mu$ and sample standard deviation $\sigma$
are  also shown. } 
\label{fig:hist}
\end{figure}

{Figure~\ref{fig:hist}} shows a histogram of the  estimated power
law index $\alpha$. The distribution, we find, is sharply 
peaked in the range $\alpha = -1.9$ to $-1.5$. The sample mean and the
sample standard deviations are $-1.3$ and $0.5$ respectively. It is
interesting that $9$ out  of $18$ galaxies in our sample have $\alpha$
in the range  $\alpha = -1.9$ to $-1.5$
 irrespective of the finer details of  their
morphology. This indicates the possibility of a common origin for 
these scale free structures. 
Note that the power law index $\alpha$ for the galaxies
NGC~925, NGC~2403, NGC~3031, NGC~3198, NGC~3621, NGC~4736 and NGC~5457
are not in the range $-1.9$ to $-1.5$. We have visually investigated
the column density maps of these galaxies to see if the difference 
in slope  is associated with some feature in the
galaxy's morphology.  
We find  that the galaxy NGC~4736 has  a \HI ring
near  its center which may cause a complicated window
function. However, none of the other galaxies   have 
such  structures which may cause a steepening or shallowing of the
power spectrum. Note that $R/r_{25}$ spans from $0.02$ to $0.3$ for
the galaxy NGC~5457, and we are actually probing 
length-scales which are quite small in comparison to the galaxy's
optical disk. The interpretation in terms of 3D turbulence for this
galaxy is therefore quite well justified.   
We may expect to observe the transition to 2D at the smallest
baselines, however  
large error-bars and the limited spatial extent at this end  makes it
hard to  detect this in the present analysis. 
All the galaxies in our sample  have spiral arms which can apparently
be a source of  large scale fluctuations. We have verified using
simulations that a model galaxy which has smooth  
emission, apart from large scale spiral arms, does not produce a power 
law power spectrum. Further, the  galaxies in
our sample all have different morphology, and one would expect their
power spectra to have   different slopes if the entire fluctuations in
the ISM were just a consequence of  the spiral arms.
 The fact that  a large fraction  of the galaxies in our sample have a
 similar value of  the power law index  makes it  unlikely that   
the measured power spectrum corresponds to the fluctuations due to the
spiral arms. 

\begin{figure}
\begin{center}
\epsfig{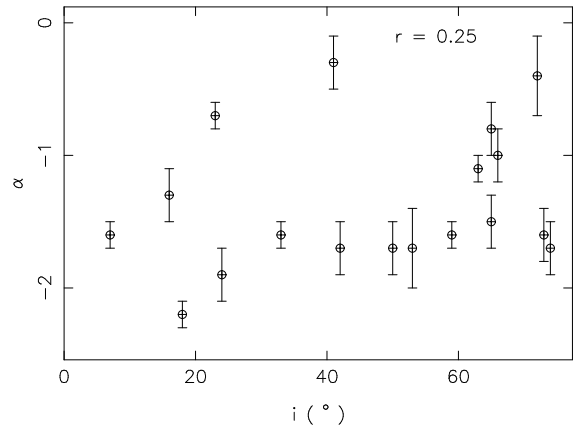}
\end{center}
\caption{Scatter plot of the average inclination angle $i$ with power law index
  $\alpha$. The $1-\sigma$ error bars of $\alpha$ is also
  shown. The value of the linear correlation coefficient $r$ is
  given at the top right corner.} 
\label{fig:inc}
\end{figure}

 The galaxies in our sample have
a wide range of inclination angles 
`$i$', ranging from $7^{\circ}$ for NGC~628 to $74^{\circ}$ for
NGC~2841. In paper IV we have performed simulations to
assess the effect of the inclination angle  on
power spectrum estimation. We found that though the length scale
range over which the power spectrum can be estimated depends on the
inclination angles, 
the power law slope remains constant irrespective of the inclination angle. 
In order to investigate if  the measured slope is influenced by the galaxy's 
inclination angle, we have evaluated  the linear 
correlation between the power law  index $\alpha$ and the inclination 
angle of the galaxy ( Figure~\ref{fig:inc}). The lack of
correlation (correlation coefficient 0.25) confirms the expectations
from the simulations  that the dispersion in the values of $\alpha$ is
not influenced by the inclination. 

It is important to identify  the mechanism that generates  the
large-scale density 
fluctuations observed in the ISM.  Here we consider a few
possibilities and  
try to assess if these mechanisms  can  give rise to the  $\alpha$
values  measured in our  sample.

\begin{figure*}
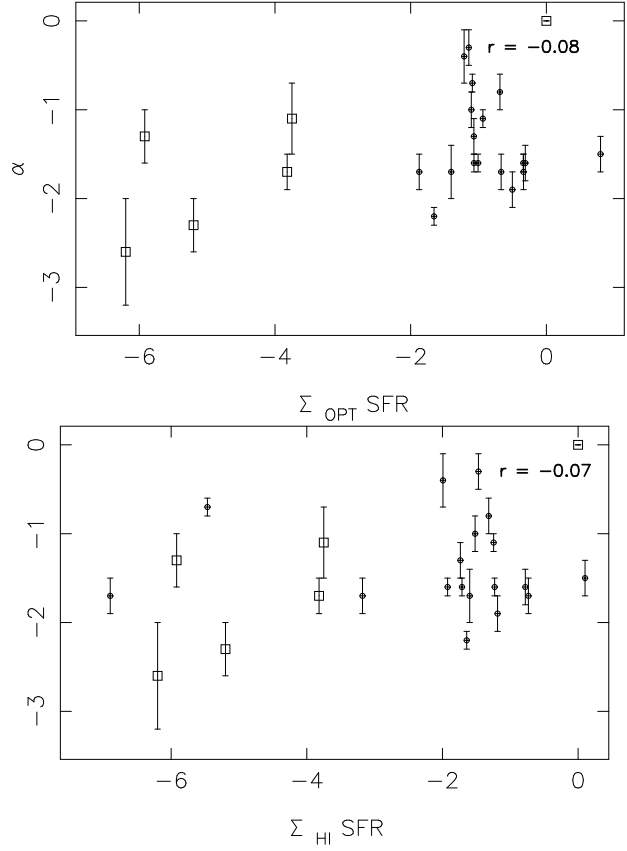

\begin{center}
\epsfig{file=SFRO.eps,width=2.2in, angle=-90}
\epsfig{file=SFRR.eps,width=2.2in, angle=-90}
\end{center}
\caption{Scatter plot of the surface density of the star formation
  rate  (Table~\ref{tab:sample})  with power law index 
  $\alpha$.  Open circles show the values from this
  paper. We also show the same quantities for the five dwarf galaxies
  from   \citet{2009MNRAS.398..887D} using open boxes.  The $1\sigma$
  error bars of   $\alpha$ are also 
  shown. The left  panel shows the SFR per unit area of the optical
  disk of the galaxies. The right panel shows similar
  quantities with the surface density of SFR calculated  per
  unit HI   disk area. We also evaluate the linear correlation
  coefficients for each of the two cases using the galaxies in this
  paper. These values are
  given at the top right corner of the each panel. The x axis of each
  plot has a unit of kpc$^{-2}$, while the SFR values are considered
  to be in units of M$_{\odot}\, $yr$^{-1}$.}
\label{fig:sfr}
\end{figure*}

At small scales,  the energy input from star formation processes
is a major driving force of the  ISM turbulence.
Star formation process 
stir up the ISM and the fluctuations hence generated diffuse to 
larger scales. If the decay process is scale independent, it can give
rise to scale free structures as seen in our study. In such a scenario,
the galaxies with higher global star formation rate would have a shallower
power spectrum. For a sample of irregular
galaxies, \citet{2005AJ....129.2186W} find that on scales of $10-400$ pc
the power law index of the  H$\alpha$ power spectra becomes steeper as 
the star formation rate (SFR) per unit area increases.  In paper IV we
reported a 
weak correlation between the SFR and the power law index  
of the \HI intensity fluctuations in dwarf
galaxies.   To investigate if
star formation is driving the turbulence we see here, we 
calculate the  linear correlation coefficient of $\alpha$ with the
disk averaged SFR per unit area of  the  optical disk as well as the
SFR per unit area of the \HI disk. The result, shown in 
Figure~\ref{fig:sfr}, suggests that    there is no correlation
between the SFR and the power law spectral   index  at the
length-scales probed in our study. 

ISM turbulence can also be driven by the kinetic energy of the \HI gas. 
In this case, one would expect the \HI velocity dispersion
$\langle \sigma \rangle$ to  be correlated with
$\alpha$. \citet{2009AJ....137.4424T} have estimated $\langle
\sigma \rangle$ as well as $\sigma_{25}$ for 11 galaxies in the
THINGS sample. We consider here the galaxies with inclination angle
smaller than $50^{\circ}$ to reduce the effect of the contribution from
large scale velocity gradients  in the observed
velocity dispersion. These galaxies are listed in {
  Table~\ref{tab:vdisp}, where columns (2) to (5) give (2) $r_{25}$ in
kpc, (3) $\langle \sigma \rangle$ in km s$^{-1}$, (4) $\sigma_{25}$ in
km s$^{-1}$ and (5) $\alpha$. Note that $\sigma_{25}$
represents the velocity dispersion at the radius corresponding to
$r_{25}$, within which major star formation takes place, whereas 
$\langle \sigma \rangle$ is an average over the entire disk. The
maximum length scales probed using the power spectrum analysis
presented in this paper is comparable to the radius $r_{25}$ for these
galaxies, however, the power spectrum estimates are over the entire
\HI disk. {Figure~\ref{fig:disp}} shows the  scatter plot of
$\sigma_{25}$ and $\langle \sigma \rangle$ with $\alpha$. We find
no correlation between $\alpha$ and  $\sigma$ for these six galaxies.
Strictly speaking, the velocity dispersion is more likely to be
correlated with the amplitude of the power spectrum  rather than its 
slope. Since the estimator that we have used here is not sensitive to  
the amplitude of the power spectrum, we can not conclusively rule out
the influence of the velocity dispersion.  

\begin{figure*}
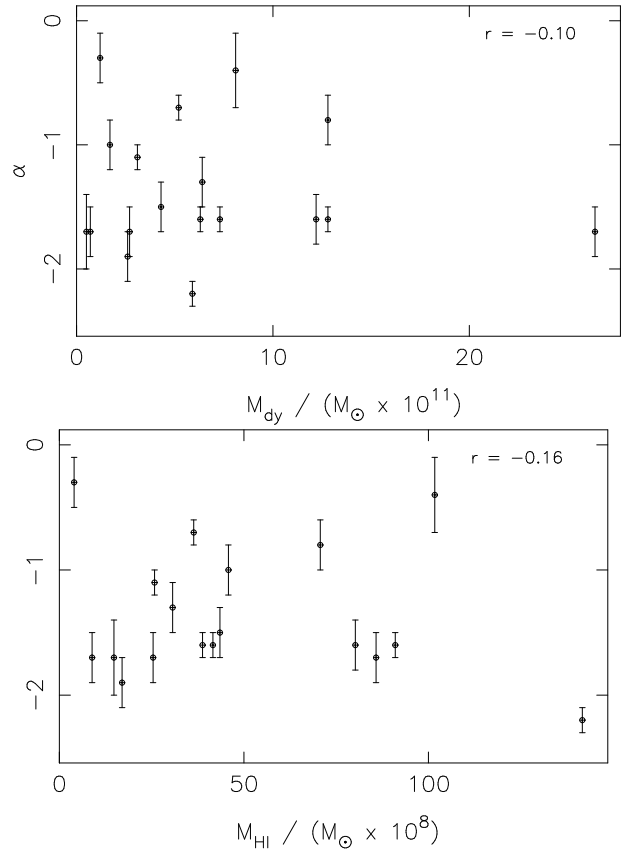

\begin{center}
\epsfig{file=MDY.eps, width=2.2in, angle=-90}
\epsfig{file=MHI.eps, width=2.2in, angle=-90}
\end{center}
\caption{Scatter plot of the dynamical mass $M_{dy}$ and the total HI
  mass $M_{HI}$ (Table~\ref{tab:sample})
  with power law index 
  $\alpha$. The $1-\sigma$ error bars of $\alpha$ is also
  shown. The value of the linear correlation coefficient $r$ is
  given at the top right corner.} 
\label{fig:ms}
\end{figure*}

It is possible that the scale-invariant power-law  power spectra
measured here may be the outcome of  the gravitational effects of  the
galaxy's dark matter halo or the self gravity of the galaxy's \HI
gas.   In  such a  scenario, it is possible that $\alpha$ is
correlated with the galaxy's  
total dynamical mass $M_{dy}$  or its  total \HI mass  $M_{HI}$.    
{Figure~\ref{fig:ms}} shows a scatter  plot of 
$\alpha$ with $M_{dy}$ (left) and $M_{HI}$  (right) respectively.  
The dynamical mass estimates are obtained either from the  
THINGS data \citep{2008AJ....136.2648D} whenever available, or from
earlier references as discussed in { Section~2}.  The $M_{HI}$
values are taken from \citet{2008AJ....136.2563W}. Our analysis yields
very low linear correlation coefficients suggesting that the gross
gravitational effects of the dark matter halo or the \HI disk are not
the responsible for the observed power-law power spectra. 

\begin{table*}
\centering
\caption{The values of  $r_{25}$, $\langle \sigma \rangle$, $\sigma_{25}$ 
for six galaxies of our sample. These values  are adopted from
\citet{2009AJ....137.4424T}.  }  
\begin{tabular}{ l r r r r }
\hline \hline
Galname   & $r_{25}$ & $\langle \sigma \rangle$ & $\sigma_{25}$ & $\alpha$ \\
          &  (kpc)  & (km s$^{-1}$)          & (km s$^{-1}$) &          \\
\hline 
 NGC~628  & $10.1$ & $ 8.0$ & $ 7.7\pm0.4$ & $-1.6\pm0.1$\\
 NGC~3184 & $11.7$ & $10.4$ & $ 9.7\pm0.9$ & $-1.3\pm0.1$\\
 NGC~4736 & $ 5.3$ & $12.0$ & $12.1\pm1.4$ & $-0.3\pm0.2$\\
 NGC~5194 & $11.4$ & $17.7$ & $17.0\pm1.4$ & $-1.7\pm0.2$\\
 NGC~6946 & $ 9.2$ & $10.1$ & $ 8.6\pm0.9$ & $-1.6\pm0.1$\\
 NGC~7793 & $ 5.7$ & $11.4$ & $ 9.7\pm1.1$ & $-1.7\pm0.2$\\
\hline
\end{tabular}
\label{tab:vdisp}
\end{table*}

\begin{figure}
\begin{center}
\epsfig{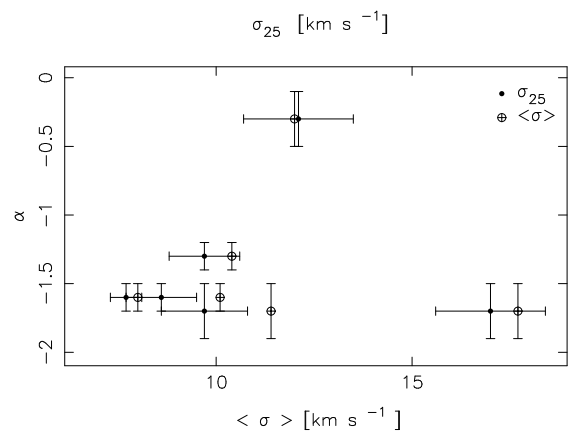}
\end{center}
\caption{Scatter plot of $\langle \sigma \rangle$ and $\sigma_{25}$ (Table~\ref{tab:vdisp})  with power law index 
  $\alpha$. } 
\label{fig:disp}
\end{figure}

In summary, the slope $\alpha$ is not correlated with any of the
galaxy parameters that we have considered. While $50 \%$ of the
$\alpha$ values are in the narrow range $-1.9$ to $-1.5$, the
remaining galaxies have $\alpha$ values spread over the broad range
$-2.2$ to $-0.3$. It is unclear if this large spread arises from
statistical  fluctuations in the distribution of $\alpha$, or if there
is  an underlying physical mechanism that is responsible for this. 

\section{Summary and Conclusion}
\label{sec:conc}
We have measured  the angular power spectrum of 21-cm specific
intensity fluctuations for a sample  of 18 spiral  
galaxies drown from THINGS.  This is the first comprehensive analysis
of the  power spectra of  a moderately sized sample of external spiral
galaxies.  For all the galaxies 
the estimated  power spectrum   can be well fitted with a single power
law across a reasonably large  dynamical  range.  The power-law power
spectrum, we find,  extends to  length-scales as large as  $\sim$ 10  
kpc indicating  the presence of scale-free structures in
the ISM on length-scales that are comparable to the size of the
galaxy.  

In this analysis we have been able to measure the \HI power spectrum over a 
large range of length-scales spanning from $\sim \ 300$ pc to
$\sim \ 16$ kpc across the entire galaxy sample.  The power spectra,
we find,  are well fit by power laws indicating the presence of
scale-invariant fluctuations.  
Fifty percent of the galaxies in our sample have a measured power law
index (slope) $\alpha$  in the range  $-1.9$ to  $-1.5$
(Figure~\ref{fig:hist}).   Only one galaxy, NGC~5457, has  a  
slope $\alpha=-2.2$ which is steeper than  $-2$. All the other galaxies 
in our sample have slopes $\alpha \ge -2$.

A large number of earlier studies (summarized in
\citealt{2004ARA&A..42..211E}),   which have probed  small
length-scales ranging from $~10$ pc to $~200$ pc, find a power 
law power spectrum with slope  $\alpha \approx-2.8$. This is believed
to be the  
outcome of three dimensional (3D) compressible turbulence in the ISM. 
If the same  process extends to  length-scales larger than the galaxy's 
scale height, we would expect to see two-dimensional turbulence in the 
plane of the galaxy's disc.   Dimensional arguments given
in Papers II and III   lead us to expect the slope of the 2D density
fluctuations to be $\alpha \approx-1.8$. This, within estimated
measurement errors,  is consistent with the slope that 
 we have measured here for most of the galaxies in our sample.  
This  prompts us to believe that the small
scales and large scale fluctuations  may both originate from the same
physical process, 
presumably turbulence. 

Energy input from supernova is believed to be the major driving
mechanism for turbulence at small length-scales. 
However, it is
unlikely that this mechanism would be able to generate large scale
coherent structures as seen here. Further, the range of length scales
that we probe here ranges from $\sim 300$ pc to $~16$ kpc which 
has practically  no  overlap with the ranges of length scale  at  which 
turbulence is believed  to be operational in our Galaxy. 

Our analysis   confirms that the power-law index  index has no
correlation with  inclination 
angle, SFR, dynamical or total \HI   mass of the galaxy.  At
present we do not have any understanding of the  physical 
process responsible for the scale-invariant large-scale  
fluctuations measured here. The  mass fluctuations  in the galaxy's
dark matter halo  is an interesting possibility that we plan to pursue 
in future.

Numerical studies of the magneto-hydrodynamic (MHD) turbulence
  \citep{2005ApJ...624L..93B,2007ApJ...658..423K,2011ApJ...736...60T,2012arXiv1205.3792B} 
  demonstrate 
that the Mach number of the medium is correlated with the  density
fluctuation power spectrum index. Alternatively, the Mach
number can also be estimated from the turbulent velocity dispersion of
the medium for such MHD turbulence. We are presently investigating
possibilities to separate the turbulent velocity dispersion in the \HI
gas from it's thermal counterpart. It may be possible that the \HI column
density as well as the velocity dispersion provides measure of similar
quantities on the turbulent \HI. However, it is not
straightforward to carry over the 
results of the simulations performed for MHD turbulence to the
turbulence in the \HI gas, which may be characteristically
different. This can be tested in future
studies and give rise to a better understanding of the ISM
turbulence in general.

\section*{Acknowledgments}

P.D. is thankful to Sk. Saiyad Ali, Kanan Datta, Prakash Sarkar,
Tapomoy Guha Sarkar, Wasim Raja and Yogesh Maan for useful
discussions. P.D. would like to acknowledge HRDG CSIR and NCRA-TIFR for
providing financial support. S.B. would like to acknowledge financial
support from BRNS, DAE through the project 2007/37/11/BRNS/357. We are
indebted to Fabian Walter  for providing us with the  \HI data from the
THINGS survey. 

\bibliographystyle{model2-names}
\bibliography{references}

\end{document}